\input ./style/arxiv-general.cfg
\documentclass[aoas,MSNbibl,nameyear,dvips]{arximspdf}
\makeatletter
   \@ifpackageloaded{graphicx}{}{\usepackage{graphicx}}
\makeatother

\doi{10.1214/15-AOAS875}
\volume{9}
\issue{4}
\pubyear{2015}
\firstpage{2110}
\lastpage{2132}
\docsubty{FLA}

\makeatletter
\newcommand{\eqref}[1]{(\ref{#1})}
\def\sfrac#1#2{#1/#2}
\def\vfrac#1#2{(#1)/#2}

\makeatother

\begin{document}
\begin{frontmatter}

\title{A stochastic space-time model for intermittent precipitation occurrences\thanksref{T1}}
\runtitle{A space-time model for precipitation occurrences}

\begin{aug}
\author[A]{\fnms{Ying}~\snm{Sun}\corref{}\ead[label=e1]{ying.sun@kaust.edu.sa}}
\and
\author[B]{\fnms{Michael L.}~\snm{Stein}\ead[label=e2]{stein@galton.uchicago.edu}}
\runauthor{Y. Sun and M. L. Stein}
\affiliation{King Abdullah University of Science
 and Technology and\break
University of Chicago}
\address[A]{CEMSE Division\\
King Abdullah University of Science\\
\quad and Technology\\
Thuwal 23955-6900\\
Saudi Arabia\\
\printead{e1}}
\address[B]{Department of Statistics\\
University of Chicago\\
Chicago, Illinois 60637\\
USA\\
\printead{e2}}
\end{aug}
\thankstext{T1}{Supported in part by the US National Science
Foundation Grants DMS-11-06862, 11-06974, and
11-07046, and the STATMOS research network on Statistical Methods in
Oceanic and Atmospheric Sciences.}

\received{\smonth{3} \syear{2014}}
%
\revised{\smonth{8} \syear{2015}}

%
\begin{abstract}
Modeling a precipitation field is challenging due to its intermittent and
highly scale-dependent nature. Motivated by the features of high-frequency
precipitation data from a network of rain gauges, we propose a threshold
space-time $t$ random field (tRF) model for 15-minute precipitation
occurrences. This model is constructed through a space-time Gaussian random
field (GRF) with random scaling varying along time or space and time.
It can
be viewed as a generalization of the purely spatial tRF, and has a
hierarchical representation that allows for Bayesian interpretation.
Developing appropriate tools for evaluating precipitation models is a
crucial part of the model-building process, and we focus on evaluating
whether models can produce the observed conditional dry and rain
probabilities given that some set of neighboring sites all have rain or
all have no rain. These conditional probabilities show that the
proposed space-time model has noticeable improvements in some
characteristics of joint rainfall occurrences for the data we have considered.
\end{abstract}

%
\end{frontmatter}

\section{Introduction}\label{secintro}

Because of its intermittent nature, high variability, and strong scale
dependence in space and time, precipitation poses significant
challenges for both measurement and modeling methods. Stochastic
models, or stochastic generators, for precipitation can facilitate the
understanding of its probabilistic structure, and can be used to
generate simulations as input into hydrologic and agricultural models,
such as for flooding, runoff, stream flow, and crop growth. Stochastic
models are also useful for many other precipitation-related problems,
such as estimating precipitation from a set of rain gauges or
validating satellite precipitation observations with surface
observations [\citeauthor{BelKun96} (\citeyear{BelKun96}, \citeyear{BelKun03})], and statistical downscaling
using stochastic precipitation generators [\citet{Maral10} and
\citet{Wil10}].
There is a substantial literature on stochastic modeling of
precipitation dating back to \citet{LeC61}. Earlier works also include
\citet{WayGupRod84} on
spectral theory of rainfall intensity, \citet{CoxIsh88} on spatio-temporal
modeling, and Rodriguez-Iturbe, Cox and Isham
(\citeyear{RodCoxIsh87}, \citeyear{RodCoxIsh88}) and \citet{Cow94} on
point process models for rainfall. Stochastic modeling of precipitation
continues to receive the attention of statisticians and hydrologists, for
example,
\citet{BerRafGne08} used latent Gaussian processes for short-term
mesoscale precipitation forecasting, \citet{SigKunSta12} proposed a
dynamic nonstationary spatio-temporal model for short-term prediction
of precipitation, and \citet{KleKatRaj12} considered daily
spatio-temporal precipitation simulation using latent and transformed
Gaussian processes.

One challenge in precipitation modeling is that the probability
distribution of precipitation depends on the space-time averaging scale
[\citet{KunSid07}]. Precipitation data are generally measured
as averages over space-time scales determined by the mechanism and
resolution achieved in a particular instrument. For example, satellite
observations provide a precipitation image with a spatial resolution of
the order of 1 km; rain gauge observations yield rain rate measurements
with collecting area as small as 200 $\mathrm{cm}^2$ and time resolution
as short as
1 minute, depending on the gauge. By analyzing rain rates on different
space-time averaging scales, it is easy to see that precipitation statistics
are strongly scale dependent. For example, the range of spatial
dependence for
monthly rain rates is much larger than that for hourly or daily rain rates.
Similarly, time dependence scales for area-averaged rain rates are
larger for
larger areas. To reflect the property of scale dependence, \citet{KunSid11} developed an empirical model of the space and time scaling properties
for rainfall occurrences. In addition, multifractal modeling in terms
of a
multiplicative random cascade process is a fairly popular choice among many
other methods, for describing spatial, temporal, or space-time multiscaling
[\citet{OveGup96}, \citet{MarSchLov96}]. These models tie a wide
range of
scales together by building multiplicative cascades and produce dependence
among different scales of the resulting process. From a statistical modeling
point of view, the rain rate can be treated as a stochastic field, and
it is
desirable to have a consistent space-time model to produce precipitation
features at different scales, rather than to have a separate model for each
scale. Therefore, it is important for any sensible precipitation models to
characterize the complex dependence structure precisely at small space-time
scales in order to produce the desired statistical properties at larger
scales. For example, averaging over adjacent space-time regions of zero and
nonzero rain produces a region that is rainy when viewed on a coarser scale.
To obtain such a wet or dry region through aggregation, the wet and dry spells
on the finer scale, driven by the spatio-temporal dependence, are essential.

Another challenge arises due to a particular feature of precipitation fields,
the intermittence, especially for small time scales. A mixed
distribution with
a point mass probability of zeros is often used to describe the frequent
occurrence of rainfall zeros [\citet{Bel87}]. Precipitation occurrence is an
important component in stochastic weather simulations, where other variables
of interest, such as temperature, humidity, solar radiation, and wind
speed, are
generally modeled conditional on the occurrence of precipitation.
For instance, Richardson's model [\citet{Ric81}, \citet{RicWri84}] has been prevalent in climate impact studies. It simulates
daily time
series of precipitation amount, maximum and minimum temperature, and solar
radiation conditional on precipitation occurrence. \citet{Kat96} studied the
statistical properties of a simplified version of Richardson's model
and used
the conditional models to generate climate change scenarios.
The spatio-temporal dependence in
rainfall zeros is a critical aspect of any space-time stochastic model for
precipitation. On the daily time scale, \citet{Kat77} used a Markov chain model
to describe the temporal dependence of precipitation occurrence at
individual locations, \citet{ZheKat08} and
\citet{ZheRenCla10} extended
the Markov chain model for simulations of the multisite precipitation, \citet{HugGut99} introduced a spatio-temporal model of precipitation
occurrence using hidden Markov models, and \citet{AilThoTho09}
developed a
hidden Markov model using censored Gaussian processes.

For many meteorological applications, especially flood warning and
drain\-age management, good short-term simulations of multisite
precipitation are required.
Modeling the spatio-temporal dependence is necessary to better
characterize the movement or the spatial patterns of the precipitation
over short time scales.
Although much progress has been achieved in the development of
precipitation \mbox{modeling}, the generation of multisite
precipitation sequences with realistic spatial dependence remains a
challenge even for the daily time scale.
Precipitation models in previous works are commonly developed for daily
data and mostly focus on reproducing means of the precipitation.
In this paper,
we assess model performance in terms of reproducing spatio-temporal dependence
in precipitation occurrence. In addition to the challenge of capturing the
marginal characteristics
of the rainfall distribution, the 15-minute time scale we consider here brings
extra challenges in capturing the spatio-temporal dependence, as well as
handling high-frequency data in time. We take advantage of high-quality
precipitation data from a network of research rain gauges in Virginia,
Maryland, and North Carolina that was deployed as part of the NASA Tropical
Rainfall Measuring Mission (TRMM) ground validation effort [\citet{TokBasMcD10}], and develop a consistent space-time stochastic model for
15-minute rain
rates measured by the rain gauges. The proposed model is based on a truncated
and transformed spatio-temporal non-Gaussian random field, where the
truncation determines the occurrence of precipitation, and the transformation
describes the distribution of the positive rainfall amounts. In this
paper, we focus on the statistical properties of precipitation occurrence
using models based on
considering when a continuous random field is above some cutoff,
so that strictly monotonic marginal transformations have no impact on our
model (assuming the cutoff is subject to the same transformation).

To model precipitation occurrences, a threshold random field model is a
natural choice. For example, the truncated Gaussian random field model
used by \citet{Bel87} for the rain rate $W(\mathbf{x})$ at a location
$\mathbf{x}$
over some specified time interval is defined as
\[
W(\mathbf{x})= %
\cases{ f\bigl(Z(\mathbf{x})\bigr), & \quad$Z(
\mathbf{x})>c$;\vspace*{3pt}
\cr
0,& \quad$Z(\mathbf{x})\leq c$,}
\]
where $Z(\cdot)$ is a stationary Gaussian random field with mean $0$
and\break $\operatorname{var}(Z(\mathbf{x}))=1$, $c$ is a cutoff chosen to
make the probability
of positive rainfall equal a specified value, and $f(\cdot)$ is a
positive monotonic function chosen to obtain a specified marginal
distribution, for instance, lognormal distribution, for the positive
rainfall amounts. \citet{Ste92} considered Monte Carlo methods for
prediction and inference for truncated spatial data based on this
model. \citet{BarPla92} proposed a spatio-temporal version
of the truncated and power-transformed Gaussian model, and
\citet{autokey11} considered a different transformation family. \citet{SanGue99} also considered a spatio-temporal truncated model and
used a Bayesian approach for model inference. Moreover,
\citet{HerGueSan09} studied the distribution of rainfall extremes under a
truncated model.
However, this model may not be adequate for 15-minute precipitation. Even
though consistent and accurate rain gauge data are available to
estimate such
a model on the 15-minute time scale, there are two main issues we
need to address. First, to model 15-minute rain rates, the value of $c$
usually needs to be quite large to account for the high proportion of rainfall
zeros. As a
consequence, precipitation occurrence is driven by the joint probabilities
of multivariate normal distributions exceeding a high threshold, and
these distributions may not have sufficient flexibility at high thresholds
to capture joint probabilities of occurrence accurately.
Second, since there is necessarily
temporal dependence for 15-minute rain rates, it is desirable to have a
space-time model rather than a purely spatial model to capture the
spatio-temporal dependence. Furthermore, it is also necessary to
fit the complicated model effectively and to develop meaningful
statistics and
visualization methods for the assessment of the model fitting.

In this paper, we develop a rich class of models for high-frequency
rainfall occurrence. We propose to model the 15-minute precipitation
occurrences by a threshold space-time $t$ random field (tRF) model.
This model is constructed through a space-time Gaussian random field
(GRF) with random scaling varying along time. The temporal dependence
in the scaling process is essential for producing a continuous
space-time $t$ process. The space-time tRF can be viewed as a
generalization of the purely spatial tRF, and has a hierarchical
representation that allows for Bayesian interpretation as well. It
includes the GRF model as a special case, and is particularly useful
for precipitation modeling on short time scales. The model structure is
motivated by the representation of a univariate $t$ random variable
\[
T=\frac{Z}{\sqrt{V/\nu}},
\]
where $Z$ has the standard normal distribution, $V$ has a $\chi^2$
distribution with $\nu$ degrees of freedom, and $Z$ and $V$ are
independent. The random variable $T$ has a heavier tail distribution
than $Z$ due to the random scaling $\sqrt{V/\nu}$.
Similarly, the randomness of the scaling process in the tRF also
increases the variability across realizations from the GRF,
which allows for a higher probability that realizations from the tRF
exceed the cutoff at more locations for a given time.
In our analysis of the 15-minute precipitation occurrences, we
generalize the
threshold space-time tRF model by letting the cutoff depend on
locations and
time, as well as including seasonality.
The seasonal variations in the marginal probability of
occurrence are fitted using
logistic regression on a series of harmonics of the annual frequency.

We also develop various quantitative and visual tools for evaluating
the dependence structure implied by rainfall occurrence models. It is a
challenge to capture all of the probabilistic characteristics of joint
rainfall occurrences from $n$ sites ($n>1$), since there are totally
$2^k$ possible events for $k$ sites of interest, where $k=2,\ldots,n$.
We propose to evaluate whether models can produce
the observed conditional dry and rain probabilities given the neighboring
sites have rain or no rain, then use the conditional probabilities,
along with
the marginal rainfall probabilities, to summarize the dependence
captured by
the model. The conditional probability plot is then developed to
display the
information.
For model fitting and validation, a feature-based approach is used,
where the quality of fit is
assessed graphically by comparing
a set of the conditional probabilities calculated
from simulations of the fitted models to observed conditional probabilities.
It is shown that the extra flexibility the proposed model allows
results in noticeable improvements in some characteristics of joint
rainfall occurrences for the data we have considered.

%
\begin{figure}[b]

\includegraphics{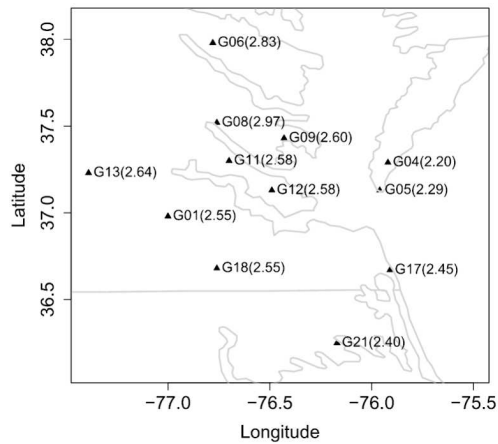}

\caption{Locations of the 12 rain gauges used in this study with the
percentage of 15-minute rainfall occurrences (in parentheses) for each
gauge site during the period of study.}
\label{figmap}
\end{figure}

The rest of our paper is organized as follows. Section~\ref{secdata}
gives a
detailed description of the rain gauge data. The dependence structure in
rainfall occurrences shown in the preliminary analyses motivates our
statistical modeling. In Section~\ref{seccompare}, we compare by
simulations the threshold Gaussian random field model to the threshold
$t$ random field model with different degrees of freedom. In
Section~\ref{subsecrf}, the purely spatial threshold $t$ random field
is introduced, and several important statistics for precipitation
occurrences are proposed under the threshold model. We then develop
useful graphical tools to display these statistics in Section~\ref{subseccpplot}. Simulation-based model comparisons are shown in
Section~\ref{subsecsim}, and the spatio-temporal threshold $t$ random
field model for precipitation occurrences is proposed in Section~\ref{subsecmodel}.
Section~\ref{secapp} presents the detailed analysis of the rain gauge
data using the proposed threshold spatio-temporal $t$ random field
model, including model inference and diagnostics. Some limitations and
possible improvements are discussed in Section~\ref{secend}.

\section{Rain gauge data}\label{secdata}

The deployment of the rain gauge network is described in detail in
\citet{TokBasMcD10} as part of the NASA Tropical Rainfall Measuring
Mission (TRMM) ground validation effort.
For quality control and reliability, each site in the network has two
or three research-quality 8-inch tipping-bucket rain gauges. These
gauges are colocated with at least one rain gauge from an operational
rainfall monitoring network.
From the 20 sites in the network, we select 12 that have essentially
complete data for the three-year period from 2004-05-19 to 2007-05-17.
The map in Figure~\ref{figmap} shows the 12 irregularly sited gauges
used in Virginia, Maryland, and North Carolina.

\begin{table}
\tabcolsep=0pt
\caption{The percentage of rainfall occurrences for different
averaging time windows from 10-minute to 91-day, where 30-day and
91-day represent the monthly and seasonal cases, respectively}\label{tabzero}
\begin{tabular*}{\textwidth}{@{\extracolsep{\fill}}lcccccccccc@{}}
\hline
\textbf{Time} &\textbf{10-min} & \textbf{15-min} & \textbf{30-min}
& \textbf{1-hr} & \textbf{3-hr} & \textbf{6-hr} & \textbf{1-day}
&\textbf{1-week}&
\textbf{30-day} & \textbf{91-day} \\
\hline
Occurrence & 1.77 & 2.55 & 4.91 & 6.47 & 10.42 & 14.77 & 32.72 & 88.57
& 99.76 &100 \\
\hline
\end{tabular*}
\end{table}

The gauges record the time of each bucket tip; one tip is equal to
0.254~mm (0.01 inches) of rain. Bucket tips are converted to rain rates
by counting the number of tips within specified intervals. We convert
bucket tips to rain rates (unit: mm/hr) within time intervals from
10-minute to 91-day.
Table~\ref{tabzero} shows the percentage of rainfall nonzeros for different
averaging time windows. For this data set, \citet{Sunetal15} used a
Mat\'ern model to describe the spatial covariance structure for
different time scales.
We can see that for shorter averaging times,
there are a large number of zeros. The fact that the 30-min frequency
is nearly double the 15-minute
frequency suggests that at least some of the 0's at the 10 and 15~minute
scales are not actually times with no rain, but intervals with not enough
rain to tip a bucket. In order to account for this effect,
we could let the cutoff for a rainfall event
change with the averaging time interval by defining, say, a rainfall event
over a 30-minute period as a period with at least two bucket tips.
However, such a definition would lead to the problematic possibility
of saying that it rained during a 15-minute interval but not over a 30-minute
interval containing the shorter interval.
Therefore, in this paper, we create a high-frequency
equally spaced time series for each gauge by considering 15-minute
averages of precipitation and
assuming no rain when there are no bucket tips in the interval.
Figure~\ref{figmap} also gives the percentage of 15-minute rainfall
occurrences for each gauge site during the period of study, which shows that
the long-term rainfall occurrence is relatively constant across the network,
although there is a hint of less frequent rainfall occurrences in the southern
part of the region and at stations G04 and G05 on the Delmarva Peninsula.
%

\section{Model comparisons}\label{seccompare}

\subsection{Truncated $t$ random fields}\label{subsecrf}

R{\o}islien and Omre (\citeyear{RisOmr06}) defined a
$t$-distributed random field (tRF) model as an extension of Gaussian random
fields (GRF) that allows for heavy-tailed marginal distributions.
On a domain $\mathcal D\subset\mathbb{R}^d$, for
$\mathbf{x},\mathbf{x}'\in\mathcal D$, the tRF is specified by its
mean function
$\mu(\mathbf{x})$,
positive definite scale function $\kappa(\mathbf{x},\mathbf{x}')$,
and the degrees of
freedom $\nu$. When the data are observed from a stationary and
isotropic tRF, $Y$, on a domain $\mathcal D$, we denote by $\kappa(h)$
the scale function between any two observations whose locations are
apart by a distance $h$. Then, the random vector $\mathbf
{Y}=(Y_1,\ldots
,Y_p)^{\mathrm{T}}$ follows a multivariate $t$ distribution, with the density
of the form
%
\begin{equation}
\label{eqpdf} \hspace*{6pt}f(\mathbf{y})=\frac{\Gamma(\vfrac{\nu+p}{2})}{\Gamma(\sfrac
{\nu}{2})(\nu\pi)^{p/2}}|\bolds{\Omega}|^{-\sfrac{1}{2}}
\biggl[1+\frac
{1}{\nu}(\mathbf{y}-\bolds{\mu})^{\mathrm{T}}\bolds{
\Omega}^{-1}(\mathbf {y}-\bolds{\mu} ) \biggr]^{-\vfrac{\nu+p}{2}},
\end{equation}
where $\Gamma(\cdot)$ is the gamma function, $\bolds{\mu}\in
\mathbb{R}^p$ is the
mean vector, $\nu\in\mathbb{R}_+$ is the degrees of freedom, and
$\bolds{\Omega}
\in\mathbb{R}^p\times\mathbb{R}^p$ is the scale matrix with $\Omega
_{ij}=\kappa
(h_{ij})$ and $h_{ij}=\|\mathbf{x}_i-\mathbf{x}_j\|$. Similar to the
Student-$t$
distribution, the tRF tends toward a GRF as $\nu\to\infty$. The
multivariate $t$ random vector can be represented by a multivariate
normal vector with random scaling $\mathbf{Y}=\bolds{\mu}+\mathbf{Z}/U$,
where $\mathbf{Z}$ and $\mathbf{Y}$ are random vectors of length $n$,
and $U$ is a
univariate random variable, providing common random scaling for each
element in $\mathbf{Z}$, with
$\nu U^2\sim\chi^2(\nu)$ and $\mathbf{Z}\sim N_n(\mathbf{0},\bolds
{\Omega})$.

Given $U=u$, the random vector $\mathbf{Y}$ has a multivariate normal
distribution with the covariance matrix $\bolds{\Omega}/u$. As $U$ is random,
the variability across realizations of $\mathbf{Y}$ is larger than the
cross-realization variability of $\mathbf{Z}$.
This scaling effect declines as $\nu$ increases,
and the tRF tends toward a GRF.

For the present application, it is not the heavy-tailed marginals of the
tRF that are important, rather it is how the tRF allows for a richer range
of spatial dependencies than the GRF when one considers where the random
field exceeds some cutoff.
Let $O(\mathbf{x})$ be the indicator of occurrence at location~$\mathbf{x}$:
\[
O(\mathbf{x})= %
\cases{ 1,&\quad $Y(\mathbf{x})>c$;\vspace*{3pt}
\cr
0,& \quad$Y(\mathbf{x})\leq c$, }
\]
where $Y(\cdot)$ is a zero-mean stationary and isotropic $t$ random
field on a
domain $\mathcal D\subset\mathbb{R}^d$ and $c$ is a cutoff indicating the
probability of
positive rainfall.
For $\mathbf{x}\in\mathcal D$, define the dry event, $D(\mathbf
{x})=\{Y(\mathbf{x})\leq
c \}$,
and the rain event, $R(\mathbf{x})=\{Y(\mathbf{x})> c \}$. Let
$p_D=P (D(\mathbf{x}) )$, $p_R=P (R(\mathbf{x}) )=1-p_D$,
$p_{D|D}=P (D(\mathbf{x})|D(\mathbf{x}') )$, and $p_{R|R}=
P (R(\mathbf{x})|R(\mathbf{x}') )$. Under the stationary
and iso\-tropic
assumptions, it is straightforward to compute the mean, $E \{
I_{D(\mathbf{x})} \}=p_D$, and
the correlations
\[
\operatorname{corr} \{I_{D(\mathbf{x})},I_{D(\mathbf{x}')} \}=\frac
{p_{D|D}-p_D}{1-p_D},
\qquad\operatorname{corr} \{I_{R(\mathbf
{x})},I_{R(\mathbf{x}')} \}=
\frac{p_{R|R}-(1-p_D)}{p_D},
\]
where $I(\cdot)$ is the indicator function. The three probabilities represent
the threshold model properties in terms of the features of precipitation
occurrence: $p_D$ is the marginal probability of the dry event for a given
location; $p_{D|D}$ and $p_{R|R}$ are conditional probabilities, describing
the spatial dependence in precipitation occurrences.

\subsection{Conditional probability plot}\label{subseccpplot}

Visualization methods can often highlight important features of the
data and are useful for model comparisons and diagnostics. For
precipitation occurrences, we propose the conditional probability plot
to visualize the degree of spatial dependence.

For illustration purposes, we choose the first $n=4000$ observations of
15-minute rain rates at the 12 locations from the rain gauge data set
described in Section~\ref{secdata}. For site $i$, $i=1,\ldots, 12$,
we compute the proportion of time that site $i$ has zero rain rates,
given all its $j$ nearest neighbors have no rain, denoted by $\varphi
_D(i,j)$, for $j=1,\ldots, 11$. Then, for example, $\varphi_D(i,1)$
means the site $i$ only conditions on one nearest neighbor, or $\varphi
_D(i,1)=P(\textrm{site }i \textrm{ dry}|\textrm{the nearest neighbor dry})$.
The conditional rain probability\break $\varphi_R(i,j)$ can be computed in a
similar way. To simplify the notation, we define the marginal dry
probability of site $i$ to be $p_D(i)=\varphi_D(i,0)=1-\varphi_R(i,0)$.

\begin{figure}

\includegraphics{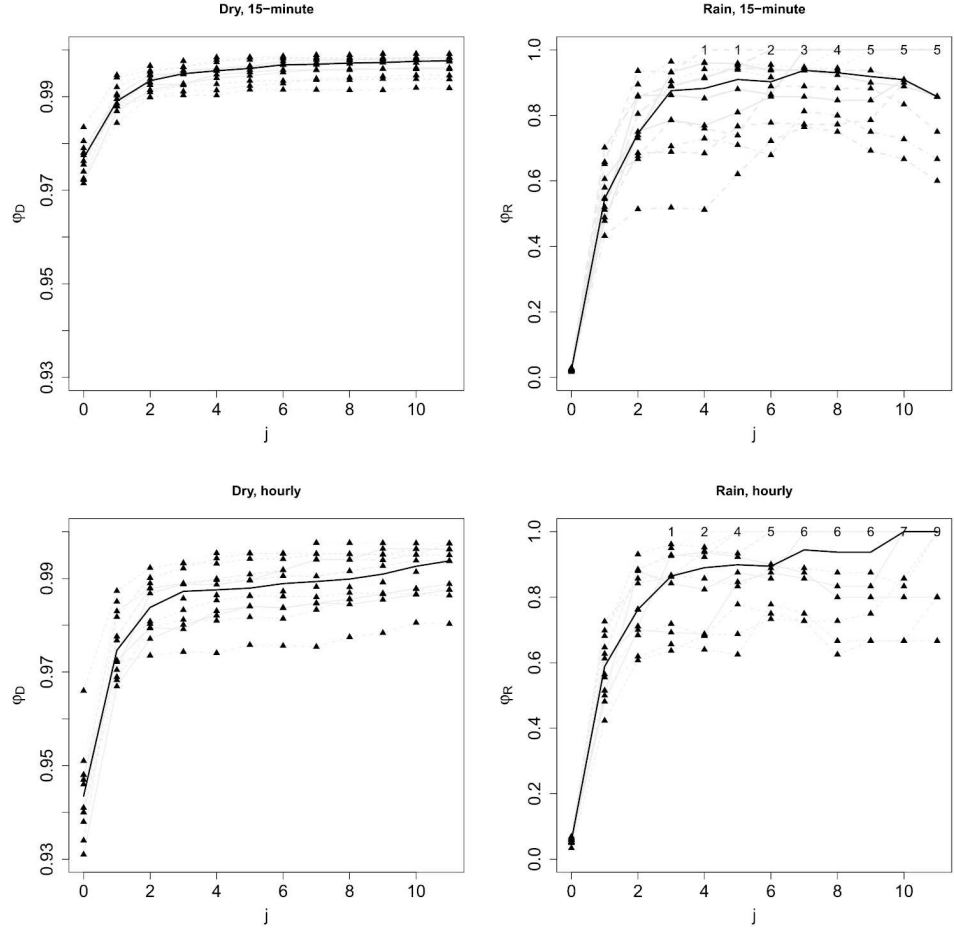}

\caption{Top panels: values of $\varphi_D(i,j)$ (left) and $\varphi
_R(i,j)$ (right) for 15-minute rain rates with $i=1,\ldots,12$ and
$j=0,\ldots,11$. Bottom panels: values of $\varphi_D(i,j)$ (left) and
$\varphi_R(i,j)$ (right) for hourly rain rates. In each figure, the
solid black line connects 12 medians at $j=0,\ldots,11$, and
probabilities from the same gauge are connected by light gray lines.
The total number of sites ($1\mbox{--}12$) for which the empirical conditional
probability is 1 is shown for a given value of $j$.}
\label{figcpplot}
\end{figure}

In Figure~\ref{figcpplot}, the top panels show the values of $\varphi_D(i,j)$
and $\varphi_R(i,j)$ for 15-minute rain rates with $i=1,\ldots,12$ and
$j=0,\ldots,11$. The bottom panels are for the cases of hourly rain rate
measurements.
Comparing the two time scales, we can see that $\varphi_D(i,j)$ is
distinctly smaller at the hourly scale than for the 15-minute scale.
In contrast, for $j>0$,
$\varphi_R(i,j)$ looks qualitatively similar at the two time scales,
although comparisons are more difficult than for dry times due to the smaller
sample sizes.

\subsection{Comparing spatial dependence}\label{subsecsim}

One way to compare and understand model properties is through multiple
simulations. In this section, we consider a purely spatial stationary
threshold $t$ random field
$Y(\mathbf{x})$ with degrees of freedom $\nu$, where $\nu=\infty$
denotes the
stationary threshold Gaussian random field. We aim to visualize the
spatial dependence implied by different models using the conditional
probability plot proposed in Section~\ref{subseccpplot}. We conduct
two simulation studies by generating independent spatial realizations
from tRF models with different $\nu$, and compare the resulting
conditional dry and rain probabilities. Since the 15-min rain rates are
necessarily correlated in time, we do not discuss the model fitting to
the real data here, but provide the detailed spatio-temporal analysis
in Section~\ref{secapp}.

\begin{figure}

\includegraphics{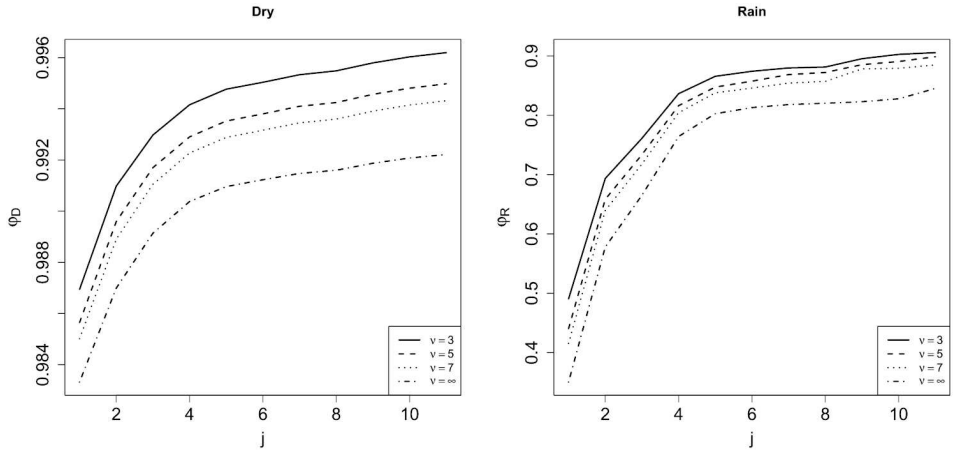}

\caption{The conditional probability plots of simulated dry (left) and
rain (right) events at 12 rain gauge sites from the threshold $t$
random field models with degrees of freedom $\nu=3,5,7,\infty$. The
marginal dry probability ($j=0$) is set to be 97.5\%. The conditional
probability is calculated for each rain gauge conditional on its $j$
nearest neighbors over 10{,}000 replications. Dashed lines in each figure
are connected medians at $j=1,\ldots,11$ as in Figure~\protect\ref
{figcpplot} for each $\nu$.}
\label{figtrfsim}
\end{figure}

First, we generate $n=10{,}000$ independent spatial fields
at the 12 rain gauge locations from a zero-mean stationary and
isotropic tRF with $\nu=3,5,7,\infty$, where the scale function has a
Mat\'ern covariance function. In
this simulation study, the Mat\'ern covariance functions with different
smoothness parameters generate similar results in terms of showing the
difference between tRF and GRF models. Here, we only present the
results from
a special case of the Mat\'ern covariance function, the Whittle covariance
function of the form
%
\begin{equation}
\label{eqwh} \kappa(h)=2\phi\alpha_0^2
\mathcal{M}_1(h/\alpha_0),
\end{equation}
where $\phi$ is the scale parameter, $\alpha_0$ is the range parameter,
and $\mathcal{M}_1=h\mathcal{K}_1(h)$ with $\mathcal{K}_1$ denoting
the modified Bessel function of order 1. We set $\phi=1$ and $\alpha
=\alpha_0/d_{\mathrm{max}}=0.5$, where $d_{\mathrm{max}}$ is the maximum
distance between the rain gauges. The cutoff $c$ is chosen to be the
97.5\% marginal quantile for each $\nu=3,5,7,\infty$, so that $p_D$
is the same for all $\nu$.
Then, the empirical values for the conditional probability of precipitation
for each rain gauge, conditional on precipitation at
its $j$ nearest neighbors, $j=1,\ldots,11$, are calculated and plotted in
Figure~\ref{figtrfsim}. This figure shows that the values of
$\varphi_D$ and $\varphi_R$ are smaller for larger values of $\nu$, the
smallest for the threshold GRF. Similar simulation studies show that
the difference between tRF and GRF is even more obvious when the cutoff
is higher. In this simulation study, the spatial correlation has the
same range for different $\nu$. One may ask whether the GRF with a
larger range parameter will be similar to the tRF. Indeed, when
computing $\varphi_R(i,j)$ for large $j$, we notice
that for data generated from the threshold GRF, there are much fewer
available conditioning sets where all the $j$ nearest neighbors have
rain, due to the low probability of exceeding a high threshold
simultaneously at many sites under the GRF. Therefore, in the second
simulation study, we allow the GRF to have a different range parameter
when compared to a tRF.

\begin{figure}

\includegraphics{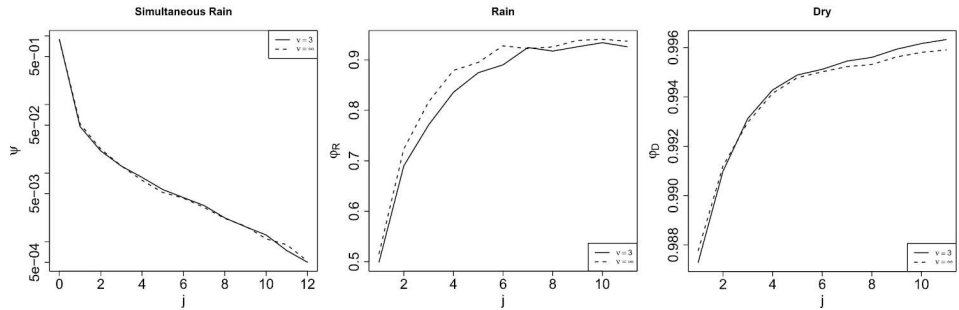}

\caption{Left panel: the simultaneous rain probabilities at exactly
$j$ sites, where $j=0,\ldots,12$, $\alpha=0.5$ for the tRF, and
$\alpha=1.055$ for the GRF. Middle panel: the conditional rain
probabilities for $j=1,\ldots,11$, where the marginal rain probability
($j=0$) is fixed at 0.025. Right panel: the conditional dry
probabilities for $j=1,\ldots,11$, where the marginal dry probability
($j=0$) is fixed at 0.975.}
\label{figfix12}
\end{figure}

Let $\psi_\nu(j)$ denote the simultaneous rain probability at exactly
$j$ sites, $j=0,\ldots,12$. For data generated from the tRF with $\nu
=3$ in the first simulation study, we compute the empirical estimates,
$\hat\psi_3(0)$ and $\hat\psi_3(12)$, respectively. For the
corresponding GRF, we numerically evaluate $\psi_\infty(0)$ and $\psi
_\infty(12)$ by the multivariate normal distribution function, and
then choose $\alpha$ such that $\psi_\infty(0)$ and $\psi_\infty
(12)$ match $\hat\psi_3(0)$ and $\hat\psi_3(12)$. Finally, we
repeat the first simulation study with $\alpha=0.5$ for $\nu=3$, and
with the selected $\alpha=1.055$ for the GRF. The conditional dry and
rain probabilities and the simultaneous rain probabilities for $\nu
=3,\infty$ are shown in Figure~\ref{figfix12}.
It is interesting that all the values of $\psi(j)$, $j=0,\ldots,12$,
are similar for $\nu=3$ and $\nu=\infty$, while the rain
probabilities of the GRF are larger than those of the tRF when
conditioning on only nearest neighbors. In other words, if exactly $j$
sites rain, it is more likely that these sites are very close to each
other in the GRF model, but for the tRF model, the $j$ sites may
contain some relatively distant ones. In fact, for the real data
application, Figures~\ref{figfbplotdm} and~\ref{figfbplotr} suggest
that the tRF model does better than the GRF model for fitting the
observed conditional probabilities because it is able to obtain lower
values for these conditional probabilities.
\subsection{Spatio-temporal model}\label{subsecmodel}
Another important aspect of precipitation occurrences is the dry or wet
spell, which is defined as the consecutive time period of no rain or
rain. Dry spells are more important and easy to define, while a rain
spell can be viewed as a sequence of consecutive time periods each with
at least one bucket tip.
To produce these statistics correctly, temporal dependence is also
important, and space-time models are then needed. Let $Z$ be a
zero-mean stationary spatio-temporal Gaussian process and $K(\mathbf{x},t)$
be the autocovariance function. For data taken regularly in time at a
modest number of sites, \citet{Ste05} proposed the following
spectral-in-time representation for $K$:
%
\begin{eqnarray}
\label{eqspcov} && K(\mathbf{x},t)=\int_{\mathbb{R}} S(\omega)C
\bigl(|\mathbf {x}|\gamma(\omega) \bigr) e^{i\mathbf{u}^{\mathrm{T}}\mathbf{x}\theta(\omega)+i\omega t}\,d\omega,
\end{eqnarray}
where $S$ is an integrable function, $C$ is an isotropic covariance
function, $\gamma$ is an even positive function, $\theta$ is an odd
function, and $\mathbf{u}$ is a unit vector. All the functions have natural
interpretations: $S$ is the temporal spectral density, $\gamma$ along
with $C$ determines the coherence at frequency $\omega$ between time
series at different locations, and $\theta$ and $\mathbf{u}$ are the phase
relationships.
\citet{Ste09} added a spatial nugget to this covariance model for
atmospheric pressure data.

We use the following parameterization for even positive functions on
$(-\pi,\pi]$ suggested by \citet{Ste05} in the covariance
function~\eqref{eqspcov}:
%
\begin{eqnarray}
\log \bigl\{\gamma(\omega) \bigr\}&=& \sum_{k=0}^La_k
\cos(k\omega ),
\\
\label{eqspcovm} \log \bigl\{S(\omega) \bigr\} &=& -\beta\log \biggl(\sin \biggl|
\frac
{1}{2}\omega \biggr| \biggr)+\sum_{k=0}^Lc_k
\cos(k\omega),
\end{eqnarray}
and choose $C$ to be a Mat\'ern covariance function with the smoothness
parameter~$\eta$, the spatial range parameter~$\alpha$, and the scale
parameter~$\phi$. The phase parameter $\theta$ is set to be 0 for
simplicity. Then $\alpha$ measures the spatial dependence at different
temporal frequencies, and $\beta$ is a long-range dependence parameter
in time.
Because of the difficulty in fitting this model, we fix $L$, the $a_k$'s,
$c_k$'s, and $C$ to values that allow good visual fits to the observed
conditional probabilities, and then vary $\alpha$ and $\beta$ to
show their effects on the process's behavior.

\begin{figure}

\includegraphics{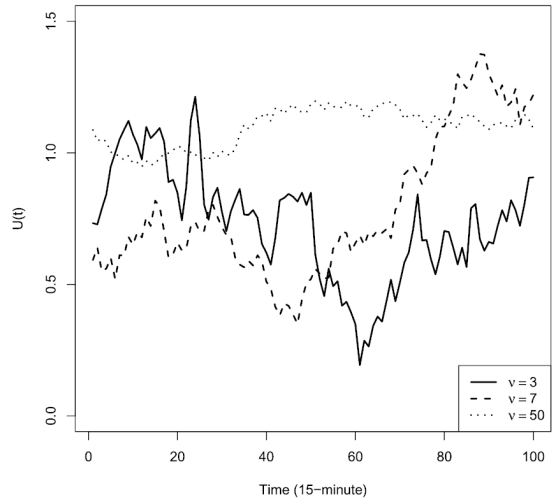}

\caption{Simulated $U(t)$ processes with $\nu=3,7,50$ and the
covariance function of $X_j(t)$ has the form of the one-dimensional
Whittle correlation function with the range parameter $\alpha_u=0.5$.}
\label{figut}
\end{figure}

Even though we introduce spatio-temporal dependence in the process, the
Gaussian random field $Z(\mathbf{x},t)$ is inadequate to characterize
the dependence
in 15-minute precipitation occurrences under the threshold model.
Motivated by
the purely spatial $t$ random field, we propose a more flexible
space-time $t$ random field model for the latent spatio-temporal process:
%
\begin{eqnarray}
\label{eqmodel} && Y(\mathbf{x},t)=\frac{Z(\mathbf{x},t)}{U(t)},
\end{eqnarray}
where $Z(\mathbf{x},t)$ is a zero-mean stationary Gaussian process,
and $\nu
U^2(t)$ is a stationary process with a margin of Gamma distribution
which can be constructed in
the following way. Let
%
\begin{eqnarray}
\label{eqden} && U^2(t)=\frac{1}{\nu}\sum
_{j=1}^{\nu}X^2_j(t),
\end{eqnarray}
where $X_j(t)$'s are i.i.d. zero-mean stationary Gaussian processes,
for $j=1,\ldots,\nu$. Then, for any given time $t=t^*$, $\nu
U^2(t^*)$ is $\chi^2_{\nu}$ distributed and it follows that
$Y(\mathbf{x}
,t^*)$ is a spatial tRF. One example of the simulated $U(t)$ process is
shown in Figure~\ref{figut}, where $\nu=3,7,50$, and the covariance
function of $X_j(t)$ has the form of the one-dimensional Whittle
correlation function given by~\eqref{eqwh} with the range parameter
$\alpha_u=\alpha_0/d_{\mathrm{max}}=0.5$.

\begin{figure}

\includegraphics{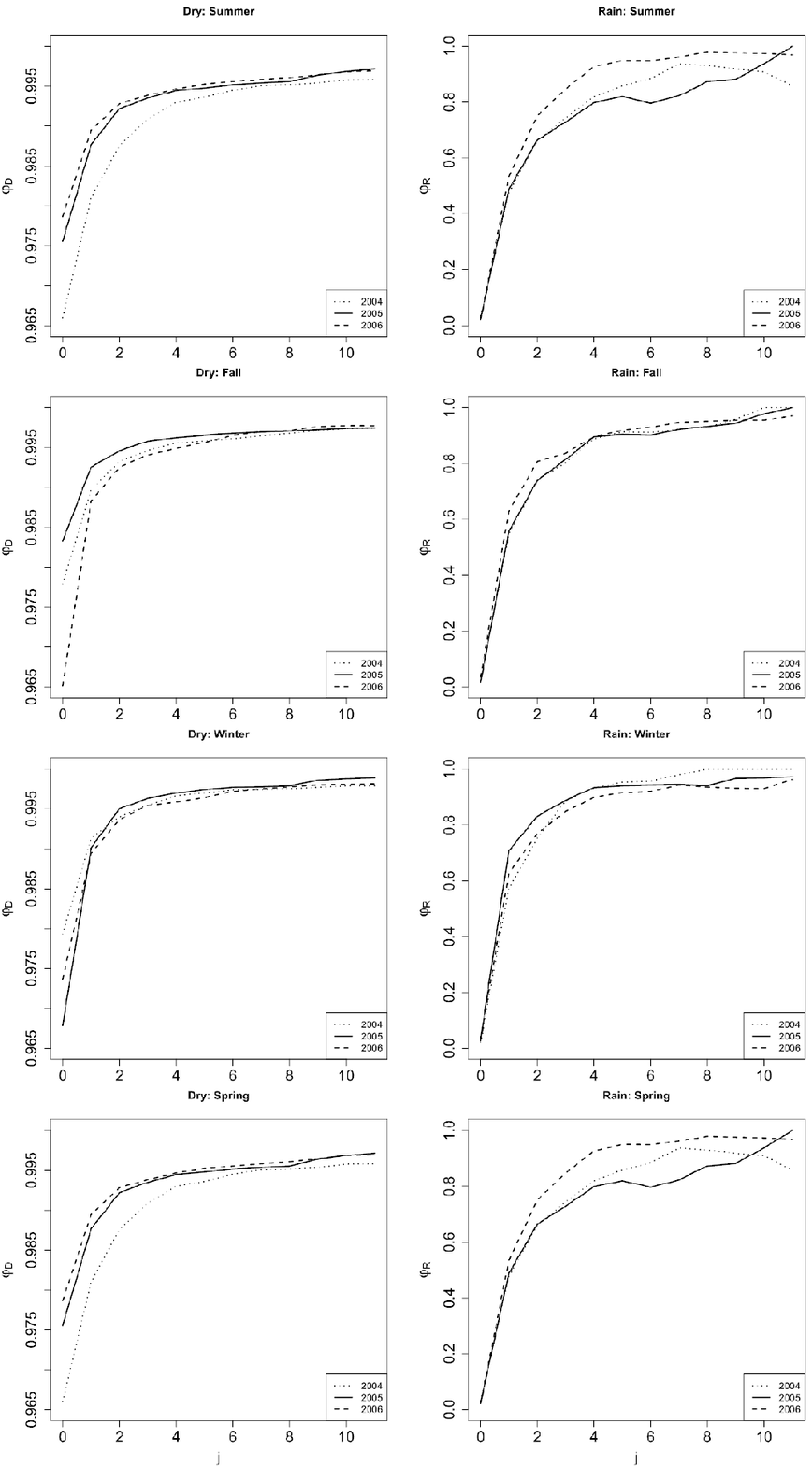}
\caption{Conditional dry and rain probabilities for the four seasons
from summer 2004 to spring 2007. Lines in each panel are the connected
medians of the 12 sites for the same season from each of the three
years.}
\label{figseason}
\end{figure}

In model~\eqref{eqmodel}, the Gaussian process $Z(\mathbf{x},t)$ is scaled
by the
process $U(t)$ randomly over time, leading to a non-Gaussian process
$Y(\mathbf{x},t)$ that increases the probability of simultaneously
exceeding a specified high quantile at many locations. The temporal
dependence in the process $U(t)$ is important in producing a continuous
non-Gaussian process $Y(\mathbf{x},t)$, because an independent $U(t)$
produces a discontinuous process that will not be adequate in general,
and taking $U(t)$ unchanging in $t$ just rescales $Z(\mathbf{x},t)$
and is
effectively no different than just changing the cutoff. Finally, the
precipitation occurrence is defined as
%
\begin{equation}
\label{eqoccur} O(\mathbf{x},t)= %
\cases{ 1,&\quad$Y(\mathbf{x},t)>c$;
\vspace*{3pt}
\cr
0,& \quad$Y(\mathbf{x},t)\leq c$,}
\end{equation}
where $c$ is a cutoff chosen to make the probability of positive
rainfall equal a specified value.

\section{Application to rain gauge data}\label{secapp}

The precipitation occurrence process
is typically nonstationary. It is location-dependent
and exhibits seasonality. Figure~\ref{figseason} shows the
conditional dry
and rain probabilities for different seasons from summer 2004 to spring 2007,
where the
four seasons are summer (June--August), fall (September--November), winter
(December--February), and spring (March--May). Lines in each panel are the
connected medians of the 12 sites for the same season from each of the three
years. The conditional probability plots summarize different patterns of
precipitation occurrences. Since 2004 and 2006 were reported
to be weak El Ni\~no years, we use 2005 as the baseline for
comparisons. We can see that the most visible interannual variability occurred
in summer. The smaller values of the conditional dry probability in summer
2004 indicate more frequent rainfall occurrences, whereas the larger
values of
the conditional rain probability in summer 2006 suggest stronger spatial
dependence of precipitation occurrences. The stronger spatial dependence
also appears in summer 2004, although it is less obvious. We can also
see that
such patterns become weaker from summer to fall in 2006. For winter and
spring, both 2004 and 2006 experience less frequent rainfall with sightly
lower conditional rain probabilities.
Different patterns of precipitation occurrences will lead to different
conditional probability curves.
For example, a process with a
small number of rainfall events of broad spatial extent could
have the same marginal rainfall probability as a process with a greater number
of localized storms, but have higher conditional rainfall probabilities given
rain at neighboring sites. Larger storms could be the reason summers
2004 and 2006 have
higher conditional rain probabilities,
since the El Ni\~no effect increases wind shear and prevents tropical
disturbances from developing into hurricanes over the Atlantic Ocean. More
detailed studies on the relationship between vertical shear and the El
Ni\~no
effect can be found in \citet{AiyTho06}. When the wind
shear is
weak, the storms grow vertically, and the latent heat from condensation is
released into the air directly above the storm, developing local
storms. When
there is stronger wind shear, the storms become more slanted and the latent
heat release is dispersed over a much larger area. Although the study region
is not typically affected by the El Ni\~no effect in terms of total
precipitation, the conditional dry and rain probabilities provide some
evidence of the different patterns of precipitation occurrence during El
Ni\~no years.

We then fit a threshold spatio-temporal tRF model to the 15-minute occurrences
for the three summers, the season for which the largest differences between
years are observed.
We let
the cutoff $c$ in~\eqref{eqoccur} depend on location and time of
year, and model
precipitation occurrence by logistic regression on a series of
harmonics to
include seasonality. Specifically, within each season of a given year, we
assume $Y(\mathbf{x},t)$ is stationary in space-time, and the precipitation
occurrence $O(\mathbf{x},t)$ is fitted using logistic regression accounting
for the
location-dependency and the hour-of-day seasonality:
\[
\operatorname{logit} \bigl[P \bigl\{O(\mathbf{x},t)=1 \bigr\} \bigr]=\alpha(
\mathbf{x} )+\sum_{j=1}^H \biggl\{
\beta_{1j}\cos \biggl(2\pi j\frac
{h(t)}{T} \biggr)+
\beta_{2j}\sin \biggl(2\pi j\frac{h(t)}{T} \biggr) \biggr\},
\]
where $h(t)\in\{1,2,\ldots,T\}$ with $T=24$ denoting the hour of time
$t$ within each day, and $\alpha$'s and $\beta$'s are coefficients.
Model fitting is conducted by the {\tt glm} function in {\tt R} [\citet{autokey25}], and the value of $H$ is chosen by AIC [\citet{Aka73}].
Then, the estimated values of the cutoff function $\hat c(\mathbf
{x},t)$ are
chosen to be the marginal quantiles corresponding to the probabilities
$1-\hat O(\mathbf{x},t)$.


Next, we need to make inference on the stationary spatio-temporal
process $Y(\mathbf{x},t)$ given the estimated cutoff function $\hat
c(\mathbf{x}
,t)$. Since model~\eqref{eqmodel} has a hierarchical representation
as the familiar Student-$t$ distribution, Bayesian methods might be
appropriate for inference on the
unknown parameters.
The EM algorithm is another natural choice, as we only observe a truncated
version of $Y(\mathbf{x},t)$.
However, these
likelihood-based methods are difficult to implement in practice in this
setting and might not be effective due to the model complexity. We
propose an
empirical approach to calibrate our stochastic model in the hope that the
model can produce statistical characteristics of the observed data. Our
estimates are obtained through the following minimization:
%
\begin{eqnarray}
\label{eqmin} && \min_{\bolds{\theta}} \Biggl[\frac{1}{M}\sum
_{k=1}^M \Biggl\{ \frac
{1}{m_1}\sum
_{i=1}^{12}\sum_{j=1}^{11}
w^D_{j}\Delta ^2_{D}(i,j)+
\frac{1}{m_2}\sum_{i=1}^{12}\sum
_{j=1}^{11}w^R_{j}
\Delta ^2_{R}(i,j) \Biggr\} \Biggr],
\end{eqnarray}
where $\Delta_D(i,j)=\varphi^{\mathrm{sim}}_D(i,j)-\varphi^{\mathrm
{obs}}_D(i,j)$,
$\Delta_R(i,j)=\varphi^{\mathrm{sim}}_R(i,j)-\varphi^{\mathrm
{obs}}_R(i,j)$,\break
$\varphi_D(i,j)$ and $\varphi_R(i,j)$ are the conditional
probabilities of the
dry and rain events defined in Section~\ref{subseccpplot},
$\varphi_D^{\mathrm{sim}}$ and $\varphi_R^{\mathrm{sim}}$ are
calculated from the
simulated data, $\varphi_D^{\mathrm{obs}}$ and $\varphi_R^{\mathrm
{obs}}$ are from
the observed data, and $M$ is the total number of simulations. Since the
conditioning set in the simulations might be empty, the conditional
probability will not be available. For the dry events, let $w_j^D$ be the
weights proportional to the number of available $\Delta_D^2(i,j)$ for each
$j$, and $m_1$ be the total number of sites, for which at least one
$\varphi_D(i,j)$ is available among $j=1,\ldots,11$. Notation for the rain
events is defined in the same way.

We generate time series with length corresponding to the number of
observations within each season of a given year, or $8736=91\times
24\times
4$ for a season with 91 days,
at the 12 rain gauge locations from model~\eqref{eqmodel} using
estimated values for all parameters.
First, we generate $u(t)$ from the scale process $U(t)$ through $\nu$
independent zero-mean stationary Gaussian processes in~\eqref{eqden},
with a Whittle covariance function, $2\alpha_u^2\mathcal
{M}_1(h/\alpha_u)$. Then, we generate a stationary space-time Gaussian
process $Z(\mathbf{x},t)$ according to~\eqref{eqspcov}--\eqref
{eqspcovm}, with $L=2$ and fixed values of $a_k$'s and $c_k$'s, and
divide it by $u(t)$. In the covariance function $K(\mathbf{x},t)$, we focus
on estimating the temporal dependence parameter $\beta$ and the
spatial range parameter $\alpha$, by fixing $\eta=1$ and $\phi=1$ in
the Mat\'ern covariance function $C$, which reduces to a Whittle
function of the form $2\alpha^2\mathcal{M}_1(h/\alpha)$. Finally,
the estimated cutoff function $\hat c(\mathbf{x},t)$ is used to
generate the
dry and rain events, $O(\mathbf{x},t)$, defined in~\eqref{eqoccur}.

The simulation procedure requires generating data from stationary
multivariate Gaussian processes in~\eqref{eqmodel} at 12 locations
and about 8736 time points. The resulting spatio-temporal covariance
matrix is of size $104{,}832\times104{,}832$. The Cholesky
decomposition of such a big matrix is difficult. Fortunately, for
multivariate regular spaced time series, the covariance matrix has a
Toeplitz structure. We apply the circulant embedding techniques in
order to use the Fast Fourier Transform (FFT) for fast and exact
simulations of stationary multivariate Gaussian time series
[\citet{WooCha94}, \citet{HelPipAbr11}].

We then estimate the set of parameters $(\alpha,\beta,\alpha_u,\nu
)$ by
minimizing the criterion~\eqref{eqmin}, where $\varphi_D^{\mathrm
{sim}}$ and
$\varphi_R^{\mathrm{sim}}$ are calculated by data generated from the threshold
$t$ random field $Y(\mathbf{x},t)$ in model~\eqref{eqmodel}.
As shown in Table~\ref{tabparest}, the consistently
small values of $\hat\nu$ for all three years suggest
the threshold tRF model fits the data better than the GRF model.
Compared to summer 2005, both summers 2004 and 2006 have smaller
estimated values of $\nu$, similar values of $\hat\alpha$, and
weaker temporal dependence estimates $\hat\beta$, although the
estimated scaling process for summer 2006 is smoother.

%
\begin{table}
\tablewidth=220pt
\caption{The estimates of $(\alpha,\beta,\alpha_u,\nu)$ in the
threshold tRF model for summer 2004, summer 2005, and summer~2006}\label{tabparest}
\begin{tabular*}{220pt}{@{\extracolsep{\fill}}lcccc@{}}
\hline
\textbf{Year} & $\bolds{\hat{\alpha}}$& $\bolds{\hat\beta}$ &
$\bolds{\hat\alpha_u}$ & $\bolds{\hat\nu}$ \\
\hline
2004&0.485& 0.486&0.199&4\\
2005&0.495&0.558&0.232&5\\
2006&0.500&0.652& 0.175&3\\
\hline
\end{tabular*}
\end{table}

\begin{figure}

\includegraphics{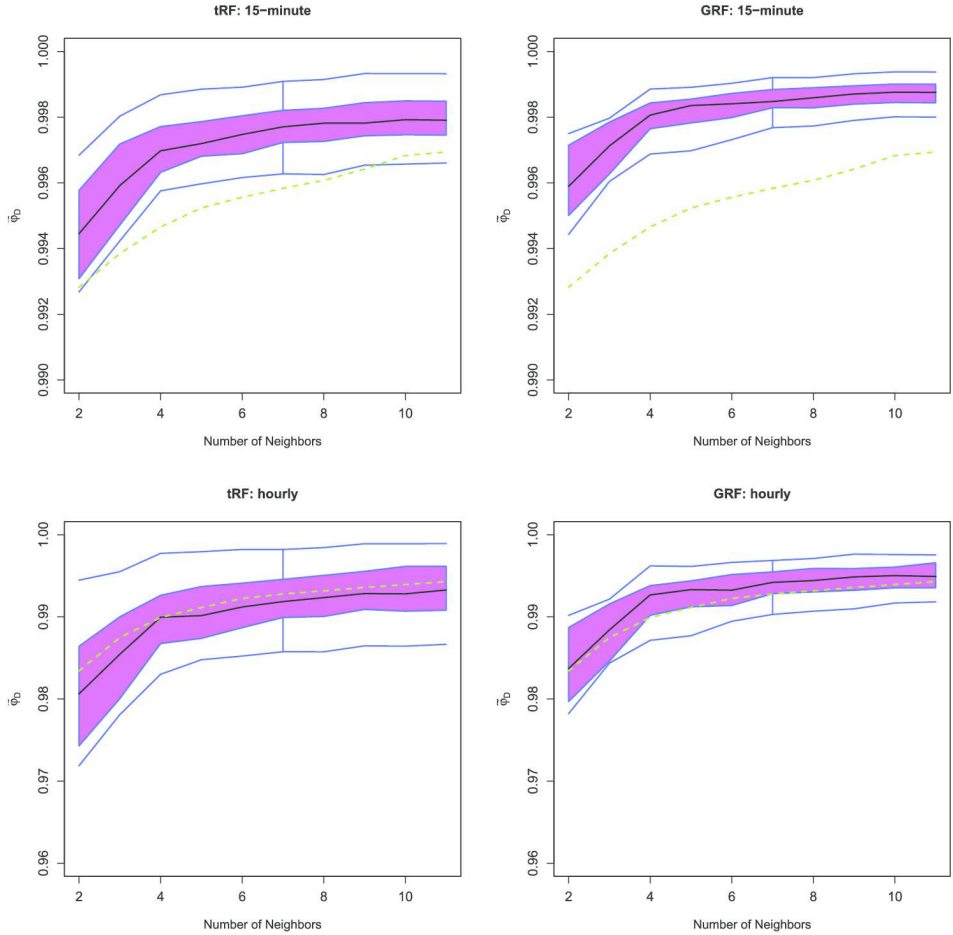}

\caption{Top panels: the functional boxplots of $\tilde\varphi
_D(j)$, $j=2,\ldots,11$, obtained from $15$-minute tRF and GRF model
simulations. Bottom panels: the functional boxplots of $\tilde\varphi
_D(j)$, $j=2,\ldots,11$, for aggregated hourly data from tRF and GRF
model simulations. In the functional boxplot, the black line is the
functional median, the middle box indicates the $50\%$ central region,
and the whiskers represent the maximum envelope of the data. The green
dashed line denotes $\tilde\varphi_D(j)$ computed from the
observations.}
\label{figfbplotdm}
\end{figure}

%
\begin{figure}

\includegraphics{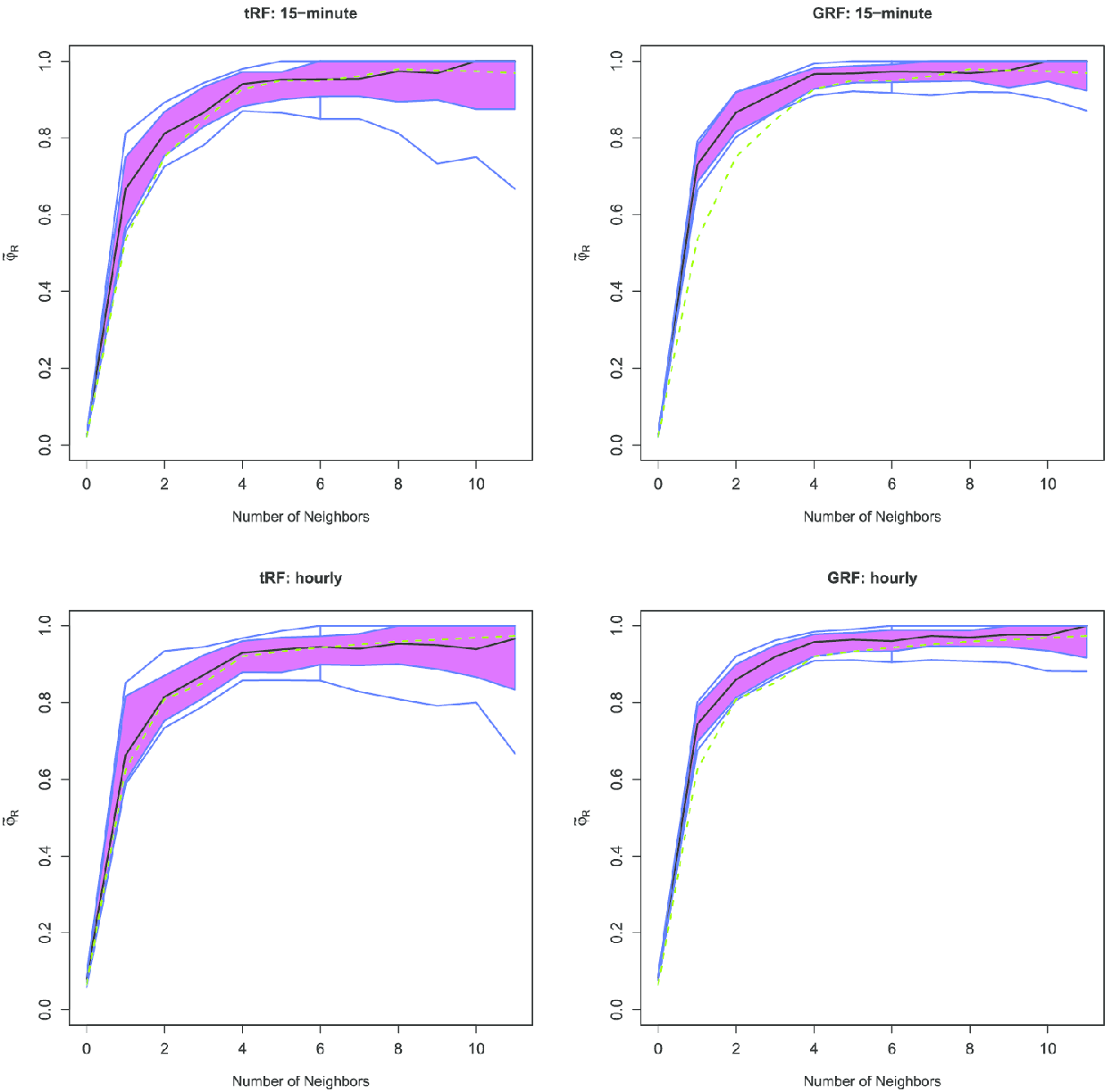}

\caption{Top panels: the functional boxplots of $\tilde\varphi
_R(j)$, $j=1,\ldots,11$, for aggregated hourly data from tRF and GRF
model simulations. Bottom panels: the functional boxplots of $\tilde
\varphi_R(j)$, $j=1,\ldots,11$, for aggregated hourly data from the
tRF and GRF model simulations.}
\label{figfbplotr}
\end{figure}

For comparisons, we also estimate parameters $(\alpha,\beta)$ in the
same way,
but $\varphi_D^{\mathrm{sim}}$ and $\varphi_R^{\mathrm{sim}}$ are
computed by data
generated from the threshold Gaussian random field $Z(\mathbf{x},t)$
in the numerator
of model~\eqref{eqmodel}. Take the data from summer 2006 as an
example. The
estimates are $(\hat\alpha,\hat\beta)=(0.811,0.123)$ for the
threshold GRF
model. The values for the minimized criterion function~\eqref{eqmin}
for the
tRF is 0.0077, and for the GRF is 0.0079. Since minimizing the
differences in the weighted conditional
probabilities in~\eqref{eqmin} is essentially fitting the model using
simultaneous rain and dry probabilities, the small
values of the criterion function for the tRF and GRF indicate that both
models fit the data well in terms of simultaneous rain and dry probabilities.
Next, we validate the fitted tRF and GRF models by comparing the
conditional probabilities of the simulated data with those of the
observed data set used to estimate the model. For each
case, we simulate  1000 seasons of
precipitation occurrences at the 12 rain gauge
locations from $Y(\mathbf{x},t)$ and $Z(\mathbf{x},t)$ in
model~\eqref{eqmodel} given
estimated parameters, and summarize the conditional probabilities of
the dry
and rain events.
Specifically, let $\tilde\varphi_D(j)$
and $\tilde\varphi_R(j)$, $j=1,\ldots,11$, be the connected medians
of the
conditional probabilities shown as solid black lines in
Figure~\ref{figcpplot}. From the simulated data, we compute 1000 such median
functions and use the functional boxplot [Sun and Genton (\citeyear{SunGen11}, \citeyear{SunGen12})] to
visualize the distribution of the conditional probability curves for
both the
generated 15-minute simulations and the aggregated hourly data, and then
compare with the conditional probability curves computed from the
observations. For the dry events, the functional boxplots of $\tilde
\varphi_D(j)$, $j=2,\ldots,11$, obtained from 15-minute tRF and GRF
model simulations are shown in the top panels of Figure~\ref{figfbplotdm}, and results for the aggregated hourly data are shown in
the bottom panels. Figure~\ref{figfbplotr} shows the functional
boxplots for the rain events.
From the functional boxplots in Figures~\ref{figfbplotdm} and~\ref
{figfbplotr}, we can see that, similar to the simulation study shown
in Figure~\ref{figfix12}, the GRF model overestimates the conditional
probabilities given rain at a moderate number of nearest neighbors,
while the tRF model can reproduce features of the observations in terms
of the conditional probabilities better.\looseness=-1

\begin{figure}

\includegraphics{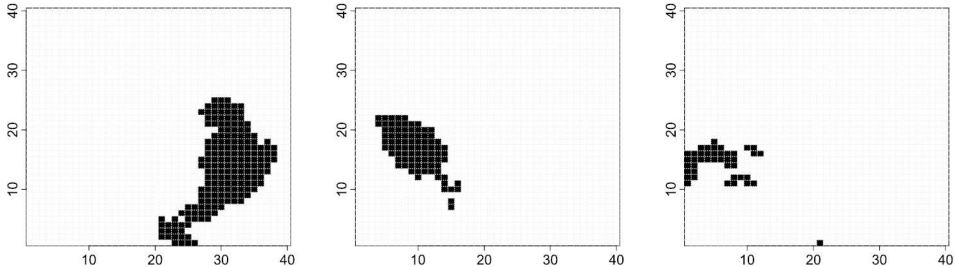}

\caption{Three examples of the simulated spatial fields for 15-minute
rain occurrences on a 40 by 40 grid.}
\label{figgrid}
\end{figure}

In the functional boxplot, the unit of information is the entire
conditional probability function. With 1000 simulations, it provides an
ordering of such conditional probability functions from the center
outward by computing the band depth values [\citet{LopRom09}]. The functional median (the black line) has the largest depth
value, representing the most central position in the sample. Then, the
50\% central region (the middle box) contains the data with the first
50\% largest depth values, and the whiskers represent the maximum
envelope of the data. The functional boxplot summarizes the
distribution of the conditional probability curves obtained from simulations.
Figure~\ref{figfbplotdm} shows that the conditional probabilities
calculated from the tRF model simulations have larger variability than
those obtained from the GRF model simulations. Consequently, the 50\%
central regions in the functional boxplots for the tRF models capture
the reality (the green dashed lines) better for the 15-minute
simulations and hourly aggregation of the dry and rain events. It
indicates that the tRF model more accurately generates the observed
conditional dry and rain probabilities.
For both the tRF and GRF models, when conditioning on a larger number
of neighbors, the variability of the conditional probability becomes
larger. However, for all the cases, the GRF model tends to produce
higher conditional probabilities compared to the observations for small
numbers of neighbors in order to achieve similar results to the
observations for larger numbers of neighbors. For the tRF model, the
conditional dry probabilities for 15-minute simulations are a little
off, but the difference in actual probability values is small. We have
also done model diagnostics for 3-hour aggregation, for which the
results (not shown) are similar to the hourly data. Overall, the tRF
models produce the observed properties well. From the fitted tRF model,
three examples of the simulated 15-minute rain occurrences on a 40 by
40 grid are shown in Figure~\ref{figgrid}.

\section{Discussion}\label{secend}

Motivated by the features of high-frequency precipitation data from a network
of rain gauges, we proposed a threshold space-time $t$ random
field (tRF) model for 15-minute precipitation occurrences. This model
has a
hierarchical representation, that is, it is constructed through a space-time
Gaussian random field (GRF) with random scaling varying along time. The
time-varying random scaling increases the variability across
realizations from
the GRF. In a threshold model for precipitation, the increased
variability is
particularly useful for small time scales, due to the lack of
flexibility of
the GRF model for high cutoff values.

We also compared the threshold GRF model to the threshold tRF models
with different degrees of freedom by simulations, and showed that the
tRF models more realistically captured dependence in 15-minute
precipitation occurrences.
We then defined several important statistics for precipitation
occurrences, and proposed useful graphical tools, the conditional
probability plot and the binary plot, to help with data visualization
and model diagnostics. The functional boxplot was used to compare model
simulations to the observations. The functional boxplot provides a way
to order functional data and display important summary statistics; it
is particularly useful to summarize functional quantities obtained from
independent simulations, and the fast algorithm developed by
\citet{SunGenNyc12} makes it more feasible in practice. For statistical inference
and model diagnostics, feature-based
approaches are used for parameter estimation and model validation.
Although the inference is not based on full likelihoods, it provides a
convenient way to reproduce features of interest, which is suitable for
applications of weather generators. For example, in the application to
the rain gauge data, we only
focused on the spatial dependence using the conditional probabilities
as the key summary statistics, and have shown that this method
effectively reproduced the spatial pattern observed in the 15-minute
rainfall occurrences. If the temporal dependence is of interest as
well, temporal summary statistics, dry and rain spells, for instance,
need to be added in the criterion for model fitting.

In this paper we have only discussed the
statistical properties of precipitation occurrence. A more complete
analysis of these data would entail using the positive rainfall amounts as
well. In principle, it would then be desirable to investigate Bayesian
inference methods under the hierarchical representation of the model,
but the
computational difficulties would be formidable. Note that it
is always possible to transform the tRF (or GRF) marginally to match any
given marginal distribution for precipitation amounts.
Indeed, if the transformation is allowed to vary in space, one can then
have a different distribution at every location.
The more critical issue, which we have not explored, is how well a
truncated and transformed tRF captures the joint distribution of precipitation
amounts at multiple sites given positive precipitation at all of the
sites or at some specified subset of the sites.
Investigation of this kind of dependence should, in our view, precede efforts
to fitting these models to the complete precipitation process
(occurrences and amounts).

Our model was developed for precipitation on short time scales and fairly
small regions. For longer scales, such as daily precipitation,
model~\eqref{eqmodel} can be modified by adding a temporal term
$V(t)$ to increase the long-term variability:
\[
Y(\mathbf{x},t)=\frac{Z(\mathbf{x},t)}{U(t)}+V(t).
\]
Here, we only aggregated to hourly and 3-hour time scales to test the
ability of the 15-minute model to aggregate realistically. In order to
obtain good fits on even longer time scales, it might be helpful to
introduce long-range dependence in $U(t)$ in model~\eqref{eqmodel}.
To handle larger regions, it will likely be inadequate to treat $U$
and/or $V$
as not depending on $\mathbf{x}$, although, to be useful,
the spatial ranges for space-time versions
of $U$ or $V$ should be much larger than the spatial range of
$Z(\mathbf{x},t)$.

\section*{Acknowledgments}
The authors thank Kenneth P. Bowman from the Department of Atmospheric
Sciences at Texas A\&M University
for providing the rain gauge data.


\printaddresses
\end{document}